\documentstyle[psfig,subfigure,lscape]{mn}

\begin{document}

\newcommand{\goodgap}{%
 \hspace{\subfigtopskip}%
 \hspace{\subfigbottomskip}}

\title[SCUBA observations of galaxies with metallicity measurements]
{SCUBA observations of galaxies with  metallicity measurements:  
 a new method for determining the relation between submm luminosity and  
dust mass}

\author[A. James et al.]{A. James, L. Dunne, S. Eales and M. G. Edmunds\\
Department of Physics $\&$ Astronomy, University of Wales, P.O. Box 913, 
Cardiff CF24 3YB, UK}

\maketitle                  

\begin{abstract}
Using a new technique we have determined a value for the constant of
proportionality between submillimetre (submm) emission and dust mass, the 
dust mass absorption coefficient ($\kappa_{d}$) at $850\mu$m. 
Our method has an advantage over previous methods in that we avoid 
assumptions about the properties of dust in the
interstellar medium.
Our only assumption is that the fraction of metals
incorporated in the dust ($\varepsilon$) in galaxies is a universal
constant. To implement our method   
we require objects that have submillimetre and far-infrared
(FIR) flux measurements as well as gas mass and metallicity 
estimates. We present data for all the galaxies with suitable
measurements, including new submm maps for five galaxies. We find
$\kappa_{850}=0.07 \pm 0.02 m^{2} kg^{-1}$.   

We have also been able to use our sample to investigate our assumption 
that $\varepsilon$ is a universal constant. We find no evidence that
$\varepsilon$ is different for dwarf and giant galaxies and show that
the scatter in $\varepsilon$ from galaxy to galaxy is apparently
quite small.

\end{abstract}

\section{Introduction}       
Measurements of the depletion of heavy elements, relative to the solar 
abundances, along the line of sight to the stars show that about $50\%$
of the metals in the Galaxy is bound up in dust grains (Whittet 1992).
This raises the questions:
Is this fraction the same for every galaxy? What are the processes that so 
efficiently lock metals into dust grains?

Although the presence of dust manifests itself in a multitude of ways - 
extinction, reddening, the polarisation of starlight, to name but three - the 
best way, in principle, to estimate the mass of dust in a galaxy is from the 
submm flux density. The advantages of this are that (a) the emission from dust
is optically thin and (b)
the emission depends mainly on the mass of dust and is only weakly
dependent on the temperature of the dust. 
The practical obstacle to making the full use of this technique has been the 
difficulty of estimating the constant of proportionality between the emission 
and the mass of dust, the mass-extinction or the mass-absorption 
coefficient. There are two methods that have been used to estimate this 
constant.

The most commonly used technique to calibrate this constant is by 
observations of Galactic reflection nebulae (Hildebrand 1983; Casey et al. 
1991). By making optical and far-infrared/submillimetre observations, one can
 estimate $A_{V}$/$\tau$$_{submm}$ in a reasonably straightforward
 way. If one assumes a value of \emph{$N_{H}/A_{V}$} and a gas-to-dust 
ratio, one can estimate the mass-absorption coefficient, $\kappa_{d}$. For 
observations at long wavelengths such as with the SCUBA submillimetre camera 
(Holland et al. 1999), which operates mostly
at 850$\mu$m, there is also the practical problem that this technique has 
only been applied at relatively short ($\lambda$ $<$ 400$\mu$m) submillimetre
wavelengths, which means that the coefficient has to be extrapolated to longer
wavelengths. Since the dust emissivity index, $\beta$, $(\kappa_{d}
\propto \nu^{\beta})$ is uncertain by a factor of two, there is an
immediate uncertainty of a factor of two in the long-wavelength value
of $\kappa_{d}$ on top of all the other uncertainties.  
This problem, of course, could
be solved by SCUBA observations of reflection nebulae, but there are 
additional problems with this technique. First, there is evidence that the
properties of dust vary with environment (e.g. Cardelli et al. 1996), and
so the dust in a reflection nebula may not be representative of a galaxy as a 
whole. Second, the technique relies on the assumption of spherical 
symmetry for the reflection nebulae and on the values for properties
that are not themselves simple to 
measure, such as the dust-to-gas ratio and \emph{$N_{H}/A_{V}$} (although
there are variations in this technique that do not require these, there are 
other assumptions that have to be made, such as the typical sizes and 
densities of dust grains). Hughes et al. (1993) estimate that the factor of
uncertainty in $\kappa_{d}(\nu)/(M_{g}/M_{d})$ at $800\mu m$ is $\sim25$. 

Alton (2000) employed Hildebrand's (1983) technique using SCUBA observations 
of an external galaxy to derive $A_{V}/\tau_{850}$, essentially using
a galaxy rather than a reflection nebula to calibrate $\kappa_{d}$. 
This method obviously avoids the problems of 
extrapolation to longer wavelengths and the fact that nebular dust may not 
be the same as dust in the rest of the interstellar medium but still requires
assumptions to be made about the physical properties of the dust.  

The other technique that has been used to estimate $\kappa_{d}$
is more loosely linked to submillimetre observations. In 
this technique one constructs a model for dust (its chemical composition, 
distribution of grain sizes etc.) that agrees with all the available data, for
example the shape of the optical extinction curve, and then uses the model to
estimate the value of the submillimetre mass-absorption coefficient (Draine \& 
Lee 1984; Hughes et al. 1993).

In this paper we suggest an alternative way of estimating the mass-absorption 
coefficient based on the global properties of galaxies rather than on 
individual properties within the Galaxy. This technique is based on the 
assumption that the fraction of metals within the interstellar medium of a 
galaxy that is bound up in dust is a constant. There is some evidence for this
from the correlation between dust-to-gas ratio and metallicity in nearby
galaxies (Issa, MacLaren \& Wolfendale 1990). The high depletion of many 
elements within our own ISM (Whittet 1992) also suggests that the mechanism
forming dust is efficient - that is, if there are metals present, they are
efficiently incorporated into dust grains. Finally, there is evidence in 
the Galaxy that although the dust extinction curve is quite variable from 
place to place, the gas phase carbon and oxygen abundances are remarkably 
constant over a wide range of ISM density (Cardelli et al. 1996; Meyer et al.
1998), implying (as these elements are the most important constituents of 
dust) that the fraction of metals in dust is a constant. In contrast to this 
evidence, Lisenfeld \& Ferrara (1998) found a non-linear relationship between 
metallicity and dust-to-gas ratio for 28 dwarf irregular galaxies, which 
suggests that $\varepsilon$, the fraction of metals locked up in dust,
is not the same for all types of galaxy. In this
paper, we will present new data that shows that Lisenfeld \& Ferrara's 
technique missed cool dust, and when this is taken into account
$\varepsilon$  does appear to be remarkably constant from galaxy to galaxy. 

To apply the technique we need a galaxy which has submillimetre and
far-infrared flux measurements, a measure of its total gass mass, and a 
measure of its metallicity.

The dust mass of the galaxy is then given by two equations:

\begin{equation}
M_{d}=M_{g}\times Z \times \varepsilon \times f
\end{equation} 
where $M_{g}$ is the mass of gas in the galaxy, $Z$ is the 
metallicity relative to solar, $\varepsilon$ is the ratio of the mass of metals in the dust
to the total mass of metals and
$f$ is the ratio of the mass of metals to the mass of gas for gas with 
solar metallicity, we estimate $f=0.019$, and;
\begin{equation}
M_{d}= \frac{S_{850} \times D^{2}}{\kappa_{850} \times B_{850}(\nu,T)}
\end{equation} 
in which $S_{850}$ is the flux density at 850$\mu$m, $D$ is the distance of
the galaxy, $B_{850}(\nu,T)$ is the value of the Planck function at 
850$\mu$m, T is the dust temperature and 
$\kappa_{850}$ is the mass-absorption coefficient at  850$\mu$m. We then set
these two equations equal to one another, and as long as we know 
$\varepsilon$, we can estimate $\kappa_{d}$.

The arrangement of this paper is as follows. In section 2 we present the 
results of a search through the archive of the James Clerk Maxwell Telescope
for submm data for galaxies with known metal abundances. In section 3 we 
derive a value for $\varepsilon$ based on observations of the local 
interstellar medium. In section 4 we apply the technique to estimate 
$\kappa_{d}$ and investigate whether $\varepsilon$ is a universal constant. 
We assume throughout that $H_{0}=75 km s^{-1} Mpc^{-1}$.

\section{Data reduction \& Results}

We looked through the JCMT\footnote{The JCMT is operated by the Joint
Astronomy Centre on behalf of the UK Particle Physics and Astronomy
Research Council, the Netherlands Organization for scientific Research 
and the Canadian National Research Council} archive for all the objects for which there 
are both abundance measurements and observations with the SCUBA camera
(Holland 1999). Table 1 is a list of galaxies in the archive for which 
there are published metallicity measurements but no published submm results.

The data for these five objects was reduced in the usual way using the SURF 
(Jenness \& Lightfoot 1998) software package. Images were first flat-fielded to
remove inhomogeneity in bolometer sensitivity. Corrections were then made for 
atmospheric absorption using the opacities ($\tau_{850}$) derived from skydip 
measurements. Noisy bolometers were masked and large spikes in the
bolometer time-stream removed by 
applying a 4-$\sigma$ clip. Residual sky noise was removed with the SURF task 
REMSKY. Following removal of the sky noise, the bolometer time-stream
was clipped again at the 3-$\sigma$ level. 

All the datasets in the archive which were taken with
the same chop throw were then made into a map using the SURF program REBIN.  
The maps were calibrated using jiggle maps of both primary and secondary 
calibrators reduced in the same way. 
Figure 1 shows contour plots of the maps of these objects overlaid on
images from the Digitised Sky Survey (DSS). The lowest level contour is $1\sigma$ and 
increasing in $1\sigma$ steps to $6\sigma$. Although it appears that
the submm emission is concentrated in the centre of NGC5253, the
dynamical range of the optical data is much greater than the submm
data, thereby making it difficult to draw any firm conclusions as to
whether the optical emission is indeed more extended than the dust emission.  

The flux was measured by using the DSS 
image as a guide to chose an aperture which contained as much of the 
galaxy and as little of the sky as possible. The flux errors were
estimated using the procedure in Dunne et al. (2000). The fluxes and
the apertures used are given in Table 1. 

Since the first draft of this paper another submm map of NGC1569 has
been published (Lisenfeld et al. 2001). The new data was taken in much better
conditions than the data we retrieved from the SCUBA archive and the
flux measured is significantly higher than ours. We repeated our
reduction  of this object but the flux determined remained $20\%$ lower 
than the flux measured by Lisenfeld et al. (2001). The discrepancy may
arise from the galaxy being larger than the SCUBA array, for which
Lisenfeld et al (2001) have tried to correct.

\begin{table*}
\(\begin{tabular}{|l|c|c|r|c|r|c|c|c|c}
\multicolumn{10}{|c|}{Table 1. SCUBA archive data and flux measurements}\\
\hline
(1)&(2)&(3)&(4)&(5)&(6)&(7)&(8)&(9)&(10)\\
Name & D &date & Run no. & int. time & $\tau_{850}$&Ap size&$S_{850}$&$S_{60}$&$S_{100}$ \\
 &(Mpc)&&&(sec)&&($\arcsec$)&(mJy)&(Jy)&(Jy)\\ \hline
UM448 &72& 13 nov 1998 & 4 & 1280 & 0.4&34&40 $\pm$9&4.1&4.3 \\
      && 13 nov 1998 & 16 & 1280 & 0.4&& \\
NGC5253  &3.2& 13 jan 1999 & 77 & 1280 & 0.4&41&192$\pm$23&30.5&29.4 \\
        && 13 jan 1999 & 114 & 1280 & 0.4&& \\
Mrk33 &24.9& 13 apr 1998 & 15 & 1280 & 0.1&36&42$\pm$8&4.7&5.3 \\
      && 13 apr 1998 & 16 & 1280 & 0.1&& \\
NGC4670 &11.0& 28 dec 1998 & 62 & 1920 & 0.5&38.6& 49$\pm$7&2.6&4.5 \\
        && 21 jan 1999 & 94 & 1920 & 0.4&& \\
        && 21 jan 1999 & 95 & 1920 & 0.4&& \\
NGC1569 &1.6& 10 sep 1997 & 81 & 1280 & 0.4&74&269$\pm32$&45.4&47.3 \\
        && 10 sep 1997 & 82 & 1280 & 0.4&& \\ \hline
\end{tabular}\)
\caption{(1) Galaxy name (2) Distance to object from Nearby Galaxies Catalogue 
(Tully 1988) (3) Date of observation (4) Telescope log run number (5)
Integration time in seconds (6) $\tau_{850}$ determined from skydips 
(7) Diameter of  aperture used to measure $850\mu$m flux (8) Measured 
$850\mu$m flux
(9) \emph{IRAS} $60\mu$m flux (10) \emph{IRAS} $100\mu$m flux}
\end{table*}

\begin{figure*}
\subfigure[NGC5253; $1\sigma=6.69\times10^{-3}Jy\: beam^{-1}$]{\psfig{file=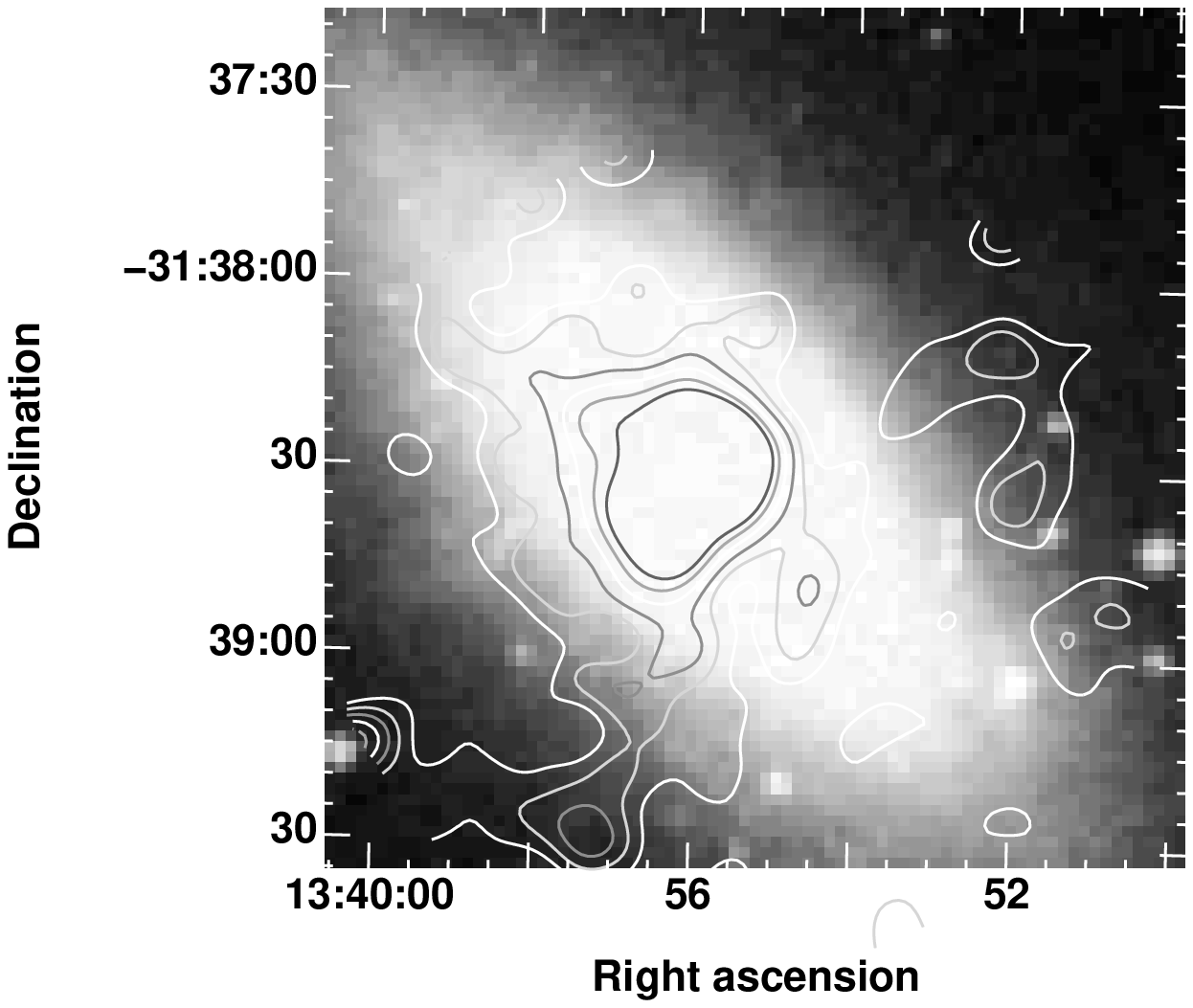,height=6.5cm,width=8.0cm}}
\subfigure[UM448; $1\sigma=4.17\times10^{-3}Jy\: beam^{-1}$]{\psfig{file=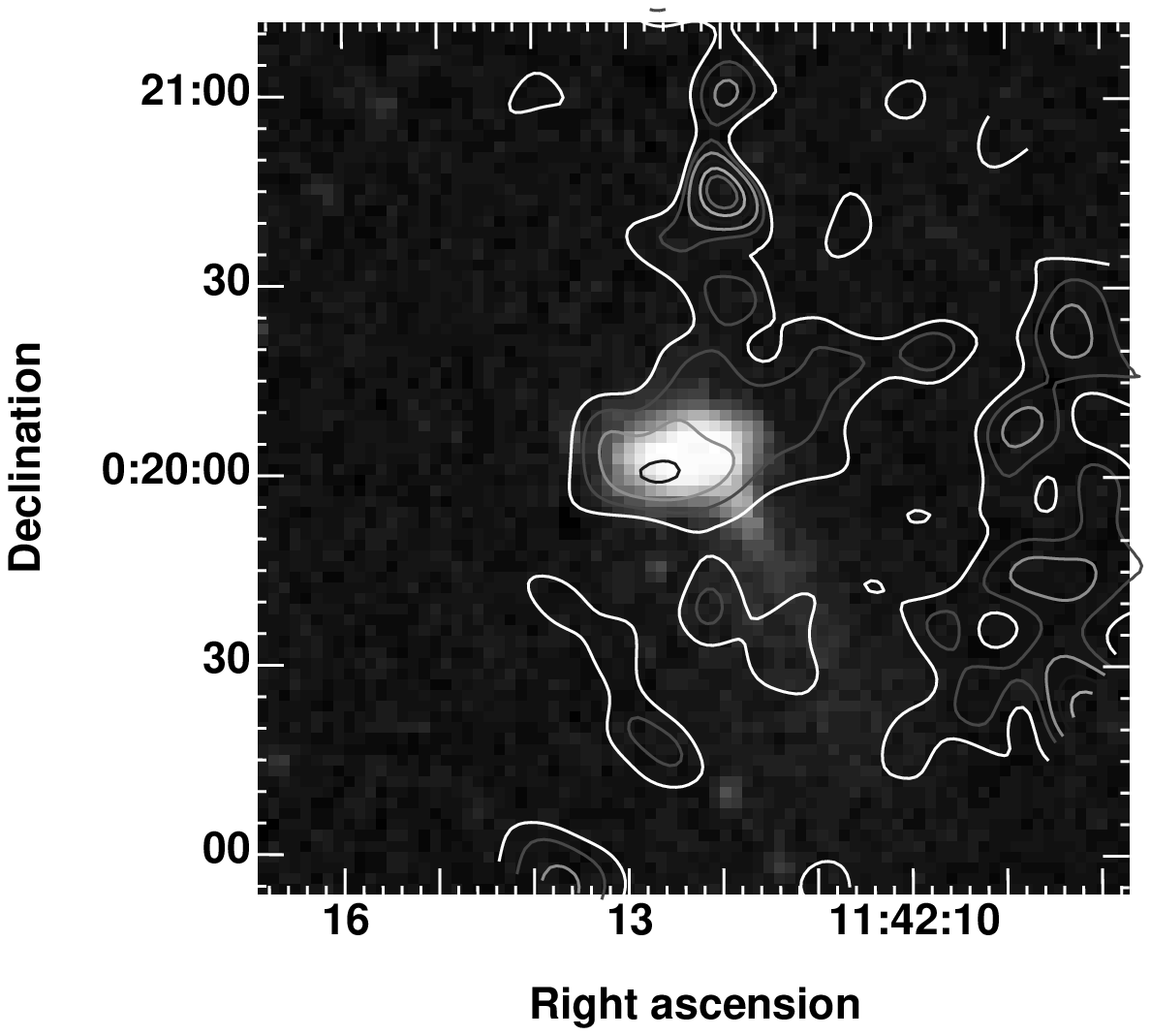,height=6.5cm,width=8.0cm}}
\subfigure[Mrk33: $1\sigma=4.23\times10^{-3}Jy\: beam^{-1}$]{\psfig{file=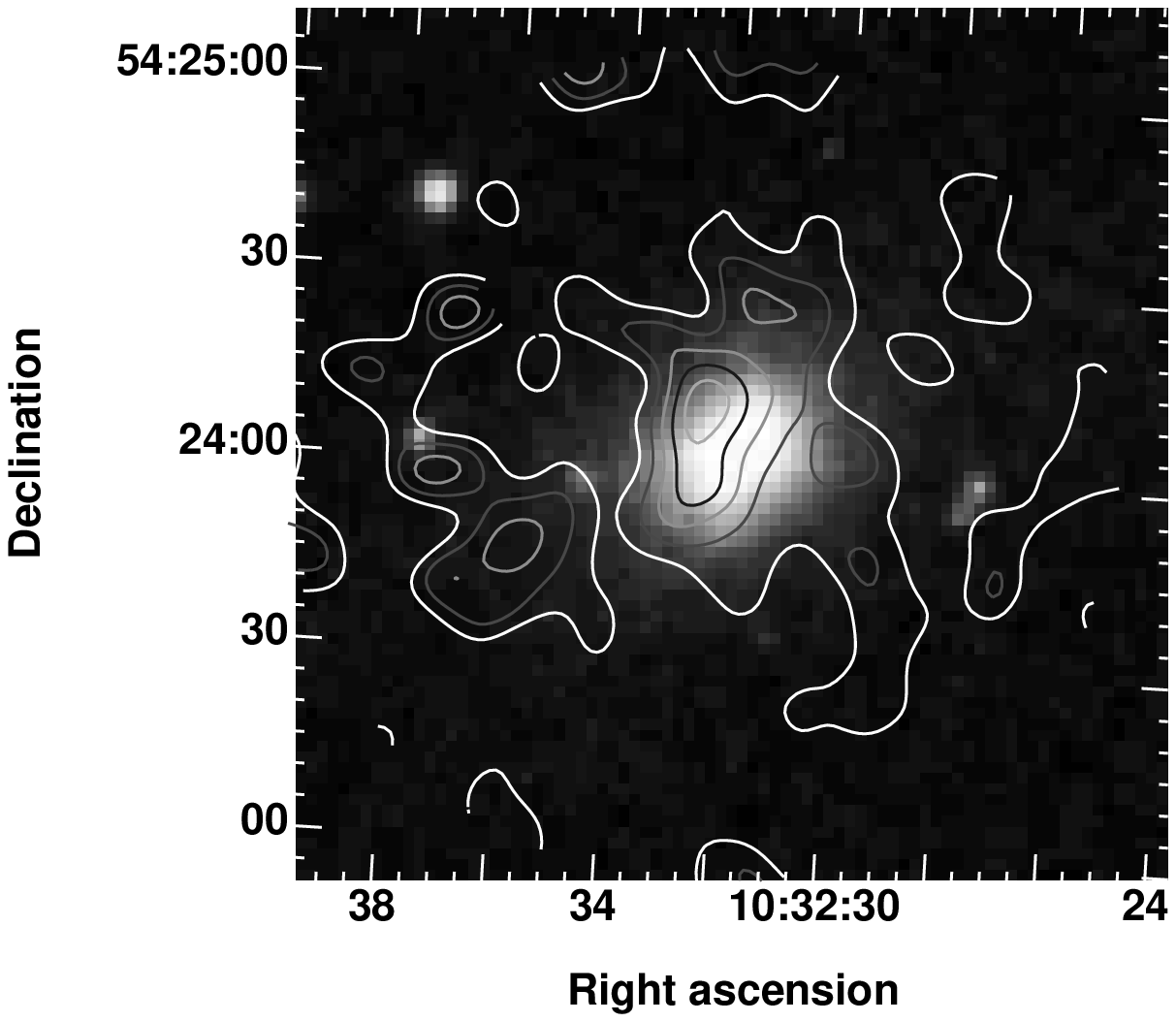,height=6.5cm,width=8.0cm}}
\subfigure[NGC1569; $1\sigma=7.32\times10^{-3}Jy\: beam^{-1}$]{\psfig{file=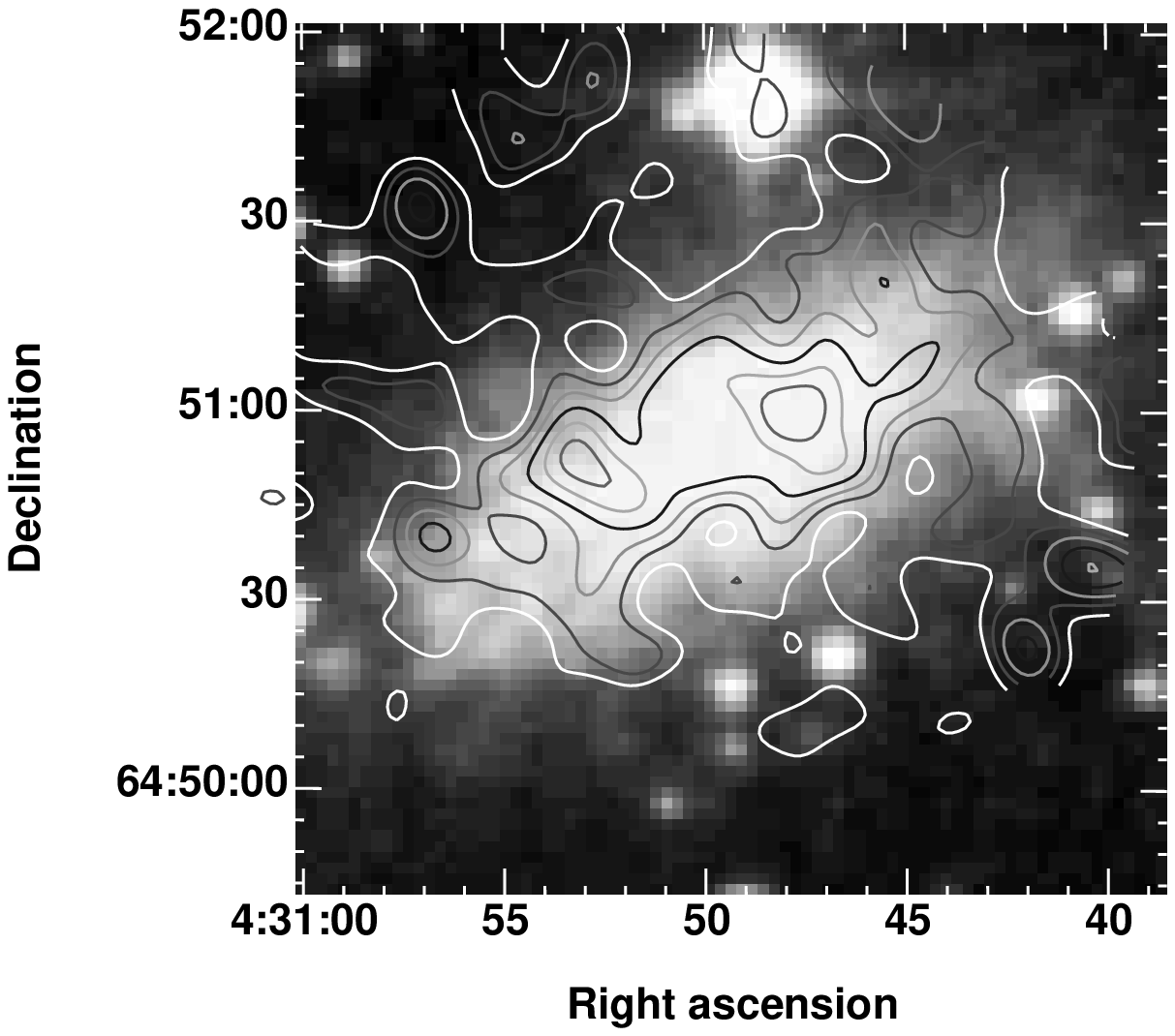,height=6.5cm,width=8.0cm}}
\subfigure[NGC4670; $1\sigma=4.24\times10^{-3}Jy\: beam^{-1}$]{\psfig{file=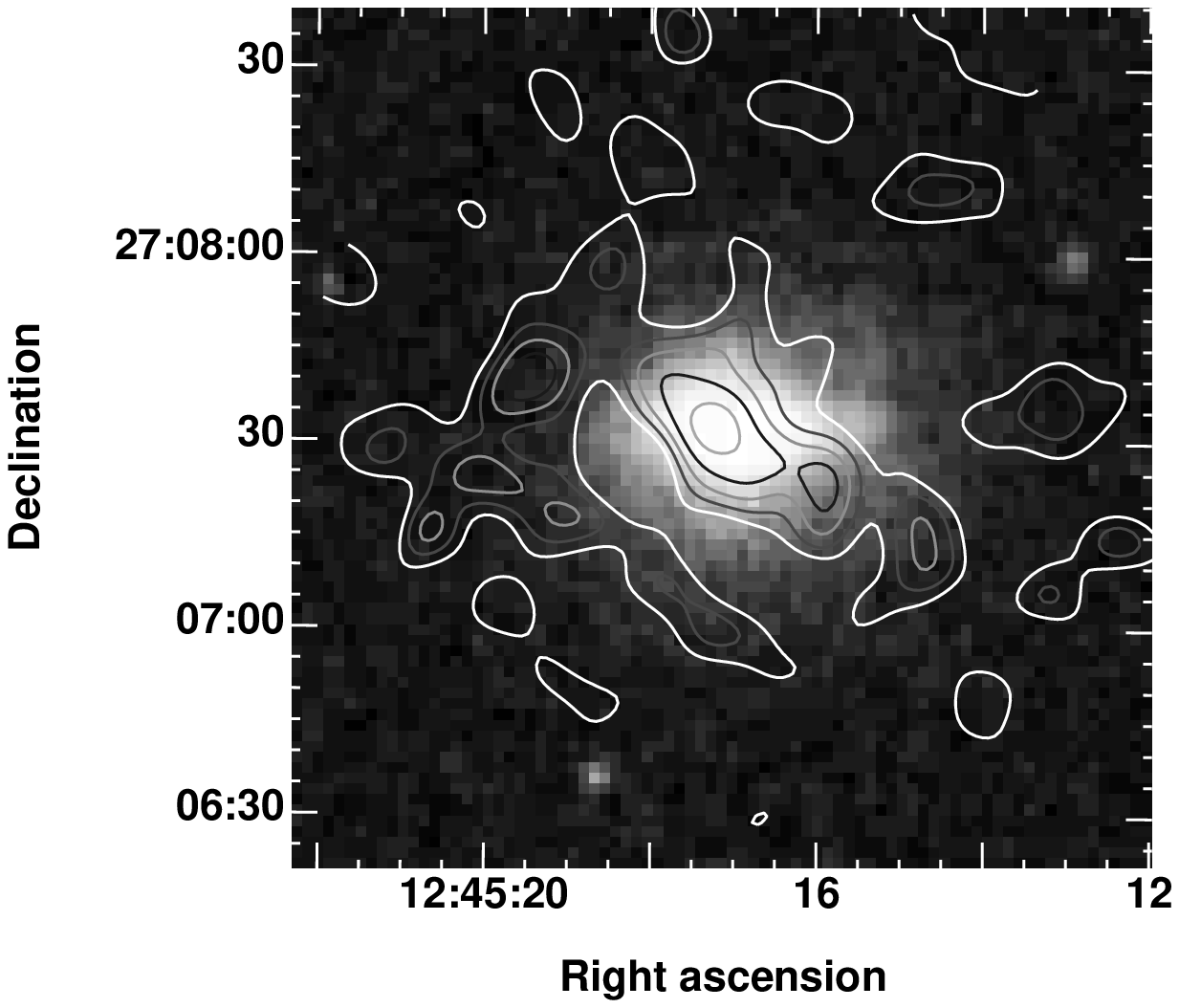,height=6.5cm,width=8.0cm}}
\caption{DSS images overlaid with SCUBA $850\mu$m contours. The
contour levels are at 1,2,3...$\sigma$}
\end{figure*}

\begin{figure*}
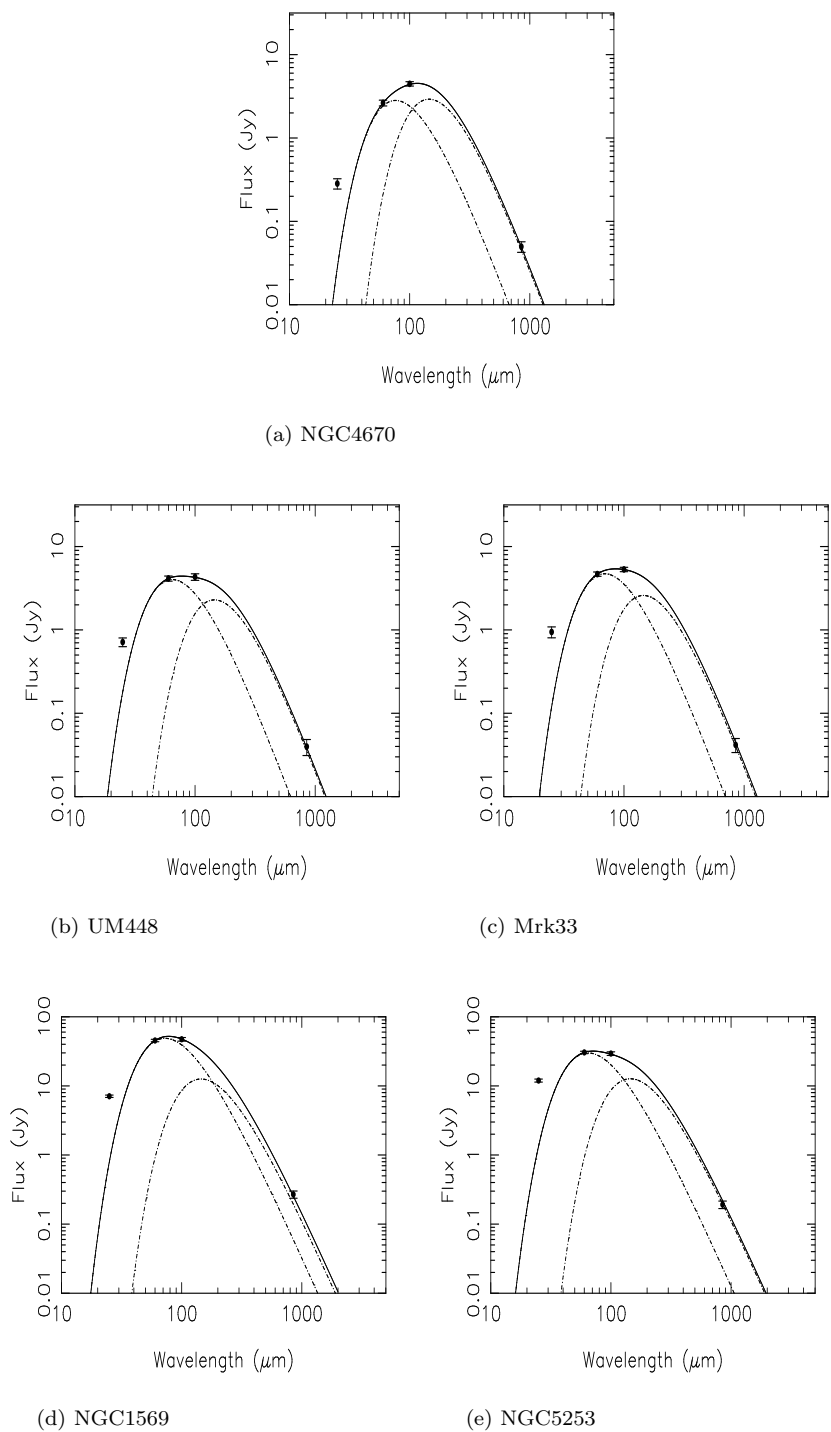

\subfigure[NGC4670]{\psfig{file=ngc4670sed.ps,height=5.0cm,width=5.0cm,angle=-90}}\\
\subfigure[UM448]{\psfig{file=um448sed.ps,height=5.0cm,width=5.0cm,angle=-90}}\goodgap
\subfigure[Mrk33]{\psfig{file=mrk33sed.ps,height=5.0cm,width=5.0cm,angle=-90}}\\
\subfigure[NGC1569]{\psfig{file=ngc1569sed.ps,height=5.0cm,width=5.0cm,angle=-90}}\goodgap
\subfigure[NGC5253]{\psfig{file=ngc5253sed.ps,height=5.0cm,width=5.0cm,angle=-90}}\goodgap
\caption{Two-component SED's assuming $\beta=2$ for the $\emph{IRAS}$ 60 and $100\mu$m 
and SCUBA $850\mu$m fluxes. The solid line represents the composite two-component SED and the 
dot-dash lines show the warm and cold components.}
\end{figure*}

\begin{table*}
\(\begin{tabular}{lcrcrc}
\multicolumn{6}{|c|}{TABLE 2. Submm results}\\
\hline
(1)&(2)&(3)&(4)&(5)&(6)\\
Name & D & $S_{850}$& $T_{w}$&$\frac{N_{c}}{N_{w}}$&log $M_{d}(S_{850},
\kappa)$\\ 
& (Mpc) & (mJy) & (K) &&($M_{\sun}$)\\ \hline 
UM448   & 72  & 40 $\pm$9& 45 &32.9 & 7.41\\ 
NGC5253 & 3.2 &192$\pm$23& 45 &24.4 & 5.39\\ 
Mrk33   &24.9 & 42$\pm$8 & 42 &22.4 & 6.51\\ 
NGC1569 & 1.6 &269$\pm32$& 40 &8.3 & 4.91\\ 
NGC4670 & 11.0& 49$\pm$7 & 38 &25.6 & 5.87\\ 
NGC2903 & 6.3 & $2120^{\dagger}$& 34 &30.7& 7.02\\
NGC4303 & 15.2& $1180^{\dagger}$& 41 &82.6& 7.54\\
NGC6946 & 5.5  &$1200^{\ddagger}$ & $\cdots$ &$\cdots$ & $7.70^{\natural}$\\  
NGC7331 & 14.3 &$1900^{\sharp}$ & 34 &44.2 & 7.69\\
NGC1222 & 32  & $84 \pm16^{\flat}$ & 38 & $6.0^{\star}$& 7.02\\
NGC7469 & 64  & $264 \pm30^{\flat}$& 38 &$12.3^{\star}$& 8.13\\
NGC7714 & 37  & $72 \pm13^{\flat}$ & 60 &$13^{\star}$& 7.05\\
NGC3994 & 41  & $106 \pm20^{\flat}$ & 33 &$9.8^{\star}$& 7.34\\
NGC3995 & 43  & $126 \pm24^{\flat}$ & 56 &$498^{\star}$& 7.47\\
NGC5929 & 33  & $119 \pm30^{\flat}$ &$\cdots$&$\cdots$& 6.51\\
NGC5953 & 26  & $182 \pm23^{\flat}$ &$\cdots$ &$\cdots$& 7.20\\
NGC5954 & 26  & $124 \pm21^{\flat}$ &$\cdots$&$\cdots$& 7.03\\
NGC6052 & 62  & $95 \pm15^{\flat}$ & 37 &$16^{\star}$& 7.65\\
MCG+02-04-025 & 122& $39 \pm18^{\flat}$ & 43 &$6.2^{\star}$& 7.83\\
IR0335+15 &138& $44 \pm9^{\flat}$ & 41 &$14.5^{\star}$ &8.01\\
NGC5256 & 109 & $82 \pm17^{\flat}$  & 36 &$8.6^{\star}$& 8.07\\
NGC7674 & 113 & $108 \pm20^{\flat}$ & 46 &$87.7^{\star}$& 8.25\\ \hline
\end{tabular}\)
\caption{(1) Galaxy name (2) Distance to object, determined using recession 
velocity taken 
from NED [The NASA/IPAC Extragalactic Database (NED) is operated by 
the Jet Propulsion Laboratory, California Institute of Technology, under 
contract with the National Aeronautics and Space Administration] 
(3) $850\mu$m flux ($\dagger$ data from Amure (priv. comm.), $\ddagger$ data from Bianchi 2000, 
$\sharp$ data from Bianchi 1998, $\flat$ data from Dunne et al. (2000)) (4) Warm component dust temperature
(5) Ratio of cold to warm dust, $\star$ results from Dunne \& Eales (2001) (6) Dust mass calculated
 using equation 4, $\natural$ dust mass taken from $200\mu$m ISO data of
Davies et al. (1999). NGC5929, 5953 and 5954
are elements of interacting pairs that were not resolved by
\emph{IRAS} but were resolved by SCUBA. To calculate $M_{d}$ for these 
objects we fitted the Dunne \& Eales two-component model to the $60,
100 \& 850\mu$m fluxes for both galaxies in the pair combined and then 
scaled the resultant dust mass using the $850\mu$m fluxes for the
individual galaxies.}
\end{table*}

\begin{table*}
\(\begin{tabular}{lccccccc}
\multicolumn{8}{|c|}{TABLE 3. Dust and gas masses using metallicity
dependent X-factors}\\
\hline
(1)&(2)&(3)&(4)&(5)&(6)&(7)&(8)\\
Name & 12 + log (O/H) &Ref. &log ($M_{HI}$) & log $(M_{H_{2}})_{A}$ &log $(M_{H_{2}})_{I}$
& log $(M_{d})_{A}$ &log $(M_{d})_{I}$\\
 & & &($M_{\sun}$) & ($M_{\sun}$)&  ($M_{\sun}$)&($M_{\sun}$) &
($M_{\sun}$)\\ \hline 
UM448   & 8.08 &1& 9.40 & 10.14 &10.92&7.36&8.08 \\ 
NGC5253 & 8.34 &1,2& 8.29 & 6.78  &7.07&5.71 &5.73\\ 
Mrk33   & 8.40 &1& 8.47 & 8.64  &8.94&6.33 &6.54\\ 
NGC1569 & 8.16 &1,3& 7.75 & 6.31  &6.92&4.99 &5.04\\ 
NGC4670 & 8.30 &1& 8.22 & 7.66  &8.10&5.69 &5.83\\ 

NGC2903 & 9.12 &4& 8.98 & 8.81 &8.03 &7.39 &7.21\\
NGC4303 & 9.01 &4& 9.42 & 9.61 &8.99 &7.90 &7.64\\
NGC6946 & 9.06 &4& 9.49 & 9.41 &8.72 &7.88 &7.69\\  
NGC7331 & 9.03 &4& 9.66 & 9.79 &9.14 & 8.13&7.87\\

NGC1222 & 8.57 &5& 9.08 & $\cdots$ &$\cdots$&6.41  &6.72\\
NGC7469 & 8.80 &6& 9.18 & 10.10 &9.80&7.99 &7.76\\
NGC7714 & 8.53 &7& 8.93 & 9.65  &9.75&7.32 &7.41\\
NGC3994 & 8.61 &8& 9.45 & $\cdots$ &$\cdots$&7.13  &7.13\\
NGC3995 & 8.66 &8& 9.79 & $\cdots$ &$\cdots$&7.52  &7.52\\
NGC5929 & 8.18 &8& 8.63 & 8.83  &9.46&6.29 &6.77\\
NGC5953 & 8.73 &8& 8.76 & 9.48  &9.29&7.62 &7.20\\
NGC5954 & 9.16 &8& 8.63 & 8.61  &7.77&7.15 &7.18\\
NGC6052 & 8.65 &8& 9.58 & 9.58  &9.71&7.60 &7.67\\
MCG+02-04-025 & 8.87 &8& 9.24 & $\cdots$ &$\cdots$&7.18 &7.18\\
IR0335+15 & 9.00 &8& $\cdots$ & 10.29 &9.69&8.36 &7.76\\
NGC5256 & 8.75 &8& $\cdots$ & 10.46 &10.24&8.28 &8.06\\
NGC7674 & 8.56 &8& 10.03 & 10.67 &10.73&8.39 &8.44\\ \hline
\end{tabular}\)
\caption{(1) Galaxy name (2) Oxygen abundance (3) References for oxygen 
abundances, 1. Lisenfeld et al. (1998) 2. Pagel et al. (1992) 3. Skillman et
al. (1989) 4. Zaritsky et al. (1994) 5. Petrosian et al. (1993) 
6. Bonatto et al. (1990) 7. Gonzalez et al. (1995) 8. see $\S$ 4
(4) HI mass (5) $H_{2}$ mass calculated using the X-factor of Arimoto
et al. (1996)  
(6) $H_{2}$ mass calculated using the X-factor of Israel (2000)  (7)
Dust mass estimated using 
eqn. 1 and the Arimoto et al. (1996) relation (8) Dust mass
estimated using eqn. 1 and the Israel (2000) relation}
\end{table*}

\begin{figure*}
\subfigure[Dust mass on the ordinate calculated using the Arimoto
X-factor (eqn. 6)]{\psfig{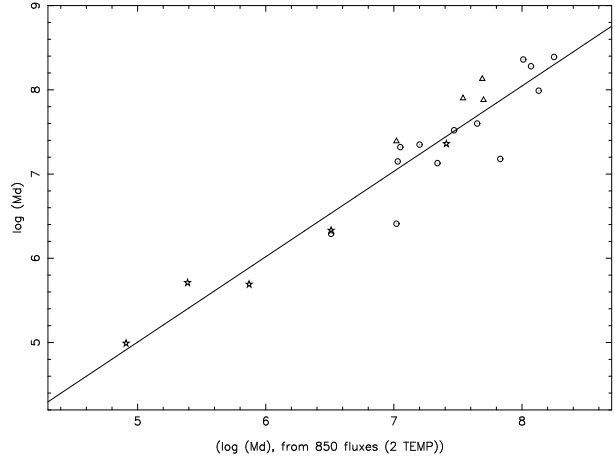}}\\
\subfigure[Dust mass on the ordinate calculated using the Israel
X-factor (eqn. 7)]{\psfig{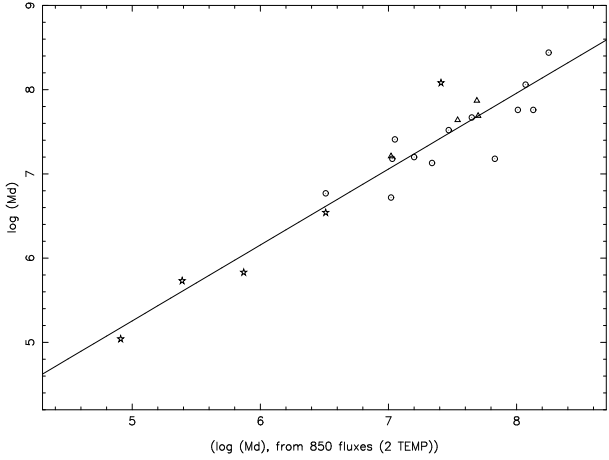}}\goodgap
\subfigure[Dust mass on the ordinate calculated using the Galactic
'standard' $X=2.8 \times 10^{20} H_{2} cm^{-2}/(k km s^{-1})$]{\psfig{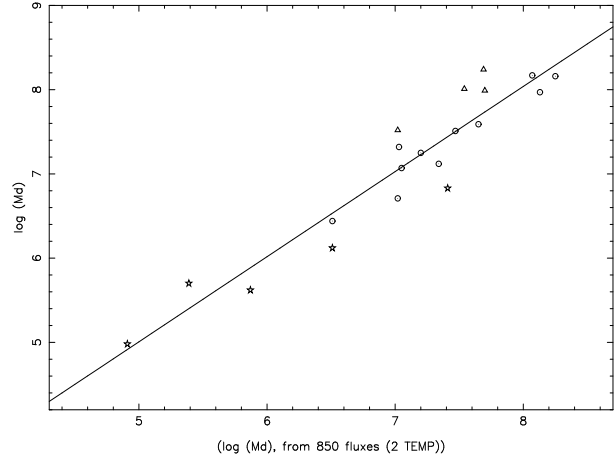}}\goodgap
\caption{Dust mass plots. The abscissa shows the dust mass calculated
using equation 4 and the ordinate displays the dust mass determined
with equation 1. The line is a least-squares best fit through the
data. The stars represent
the objects for which we reduced data from the SCUBA archive, the
circles represent the 
SLUGS sample and the triangles represent galaxies whose fluxes were obtained 
from Amure (priv. comm.) \& Bianchi (1998,2000). }
\end{figure*}

\begin{figure*}
\subfigure[Gas mass on abscissa calculated using the Arimoto X-factor
(eqn. 6). Plots using the Israel (2000) or Galactic standard X are
almost identical.]{\psfig{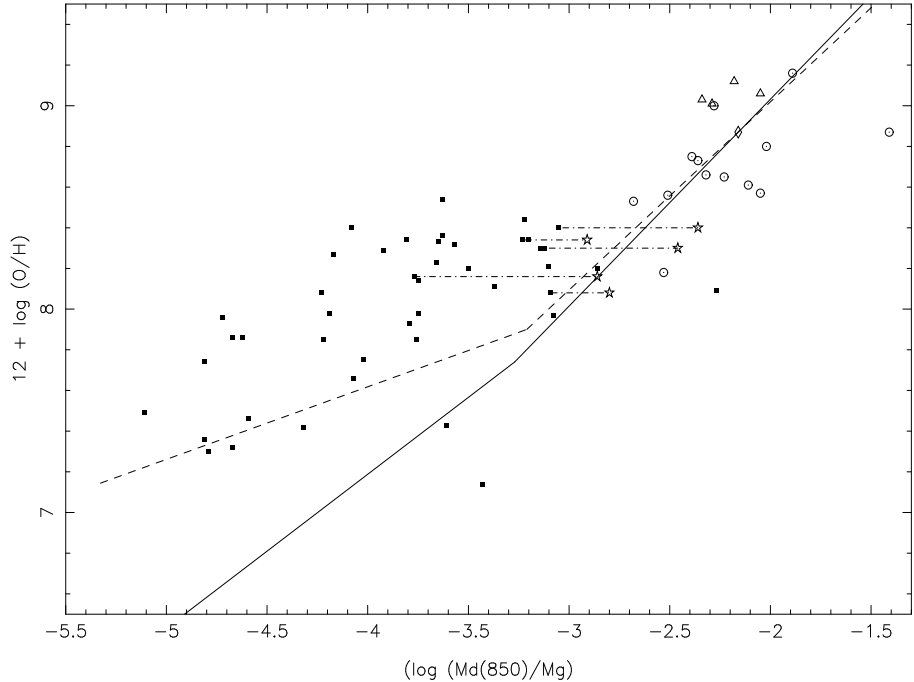}}\goodgap
\caption{Dust-to-gas ratio and oxygen abundance. The stars represent
the objects for which we took data from the SCUBA archive, the circles
show the 
SLUGS (Dunne et al. 2000) sample and the triangles are the galaxies
whose fluxes were obtained 
from Amure (priv. comm.) \& Bianchi et al. (1998,2000). The filled
squares are the data of 
Lisenfeld et al. (1998). The dot-dash lines 
indicate the distance that the Lisenfeld \& Ferrara (1998) galaxies 
move on the plot when we use our submm fluxes rather than IRAS fluxes
to estimate the dust masses.  The two lines are the predictions of 
dust-formation models from 
Edmunds (2001), which have been normalised to the Galaxy 
(the open diamond). Both models imply that a constant fraction of
metals are incorporated in dust at moderate and high metallicities. The solid line represents a model in
which supernovae play an important role in dust 
production and the dashed line a model in which dust is only produced
in some types of 
evolved low and intermediate mass stars. Some of the outlying points
on these plots lack either HI or $H_{2}$ gas masses. With gas masses
these data would move to tighten the correlation.}
\end{figure*}

\section{Calibrating $\varepsilon$}
The fraction of metals incorporated in dust in the interstellar
medium can be estimated from $UV$ spectroscopic observations of nearby stars,
which show that a significant fraction of the metals that
one would expect, from cosmic abundances, to be in the ISM
are not there - they have been bound up in dust. If $n_i$ is the
number of atoms of a given chemical element relative to
the number of hydrogen atoms, $a_i$ is the atomic weight
of the element, and $\delta_i$ is the proportion of that element 
bound up in dust, then
\begin{equation}
\varepsilon=\frac{\sum_i n_i \times \delta_i \times a_i}{\sum_i n_i \times a_i}
\end{equation}
Recent $UV$ observations of nearby stars with the Hubble
Space Telescope have made it possible to calculate $\varepsilon$ more 
accurately because they have provided accurate depletion measurements for
two of the most important elements, oxygen and carbon, strong
evidence that the depletion is the same along different sight lines,
and evidence that the cosmic metal abundance is about two thirds the
solar metal abundance (Cardelli et al. 1996; Meyer et al. 1998). 
To calculate $\varepsilon$, we have assumed the solar metal abundance
given in Pagel (1997) and then multiplied this by two thirds to
obtain the cosmic metal abundances. It should be noted that the recent 
re-determination of the solar oxygen abundance by Allende Prieto et
al. (2001), suggests that the solar metallicity is considerably closer 
to the ISM value than has been conventionally assumed. We have assumed
the  depletions
given by Whittet (1992), except that we have preferred
the recent measurements for carbon and oxygen obtained from the
Hubble Space Telescope observations (Cardelli et al. 1996; Meyer et al. 1998).
We obtain a value for $\varepsilon$ of 0.456.

\section{Applying the method}
Table 2 lists the submm results for all the galaxies that we are aware of which
have metallicity measurements and submillimetre measurements. Many of
the objects in the table are taken from the SCUBA Local Universe
Galaxy Survey (SLUGS) of Dunne et al. (2000).
Distances for objects with radial velocities $v<2000kms^{-1}$ were taken from 
the Nearby Galaxies Catalogue (Tully 1988). Table 3 gives the
metallicity measurements for the sample and also measurements of the
masses of the atomic and molecular gas.  
Metallicities for some SLUGS objects were estimated using strong line
ratios and the calibration method of Edmunds \& Pagel (1984), relating 
$R_{23}$ and $[NII]/H_{\alpha}$ and oxygen abundances in HII regions. Line
ratios were taken fron Keel et al. (1985) and Veilleux et
al. (1995). Although more recent work has shown that the Edmunds \&
Pagel calibrator may over-estimate metallicities above solar, it is
probably a reasonable estimator in the abundance range of these galaxies.

There are two complications to applying the simple method we
have outlined in the introduction: (a) the requirement
of equation (2) for a temperature model; (b) the fact that
estimates of gas mass depend to some extent on assumptions
about metallicity. We will now discuss these in turn.

\subsection{Dust Masses}
Dunne $\&$ Eales (2001), in their investigation of the FIR/submm
spectral energy distributions (SEDs) of galaxies in the 
SCUBA Local Universe Galaxy Survey (SLUGS), 
found that the 450/850$\mu$m flux ratio is
remarkably constant for galaxies with a wide range of
luminosities. They conclude that the dust emissivity index $\beta$
must be $\sim$2. They then showed that with this emissivity index,
there must be at least two dust components with different temperatures 
in galaxies, even in ultraluminous IRAS galaxies like Arp220.

We fitted the Dunne \& Eales two-component model to the five objects
for which we have new SCUBA data. Since we only
have three fluxes for each galaxy (60, 100 and $850\mu$m), we 
fixed the temperature of the cold component to be 20K, the average of
the values found by Dunne \& Eales (2001). Figure 2 shows the fits for 
the five galaxies and Table 2 gives the temperature of the warm component 
and the ratio of the cold-to-hot dust ($N_{c}/N_{w}$). We have also
given the results of the two component fits for the SLUGS galaxies by
Dunne \& Eales (2001) and values for a few other galaxies for which we 
either made 
two-component fits or for which we took two-component fits from the
literature. From the results of these fits we
have estimated dust masses using equation (4):
\begin{equation}
M_{d}= \frac{S_{850} \times D^{2}}{\kappa_{850}}\times
\left[\frac{N_{c}}{B_{850}(\nu , T_{c})}+ \frac{N_{w}}{B_{850}(\nu , T_{w})}\right]
\end{equation}
in which $S_{850}$ is the flux density at 850$\mu$m, $B_{850}(\nu,T)$ is the
value of the Planck function at 
850$\mu$m, $\kappa_{850}$ is the mass-absorption coefficient at
850$\mu$m and D is the distance found for $\Omega_{0}=1$.
We have initially adopted the value for $\kappa_{d}(850\mu$m) of 
$0.077 m^{2} kg^{-1}$ used by Dunne et al. (2000), although of course
one of the goals of the present paper is to measure a new value for
this. These initial estimates of the dust masses are given in Table 2.

\subsection{Gas Masses}
The mass of HI can be unambiguosly calculated from the intensity of
the 21-cm line. Following  Lisenfeld \& Ferrara (1998), we have
calculated the dust mass within the optical disk, since the
metallicity (and thus the dust content) outside the optical disk is
likely to be low.

Cold $H_{2}$ is detected indirectly by means of a tracer such as CO. The 
$CO-H_{2}$ conversion factor is however highly variable depending on
 temperature, density and metallicity. It is usual to determine the
$H_{2}$ mass in a galaxy by applying a 
'standard' conversion factor, $X=N(H_{2})/I(CO)$, where I(CO) is 
the velocity-integrated CO intensity and $N(H_{2})$ is the molecular 
hydrogen column density. A complication here is that the conversion
factor almost certainly depends on metallicity. We investigated the
dependence of our method on this uncertainty by trying a number of
suggested relationships between X and metallicity. 
 
Molecular hydrogen ($H_{2}$) masses for the five objects from
the SCUBA archive  were 
calculated from the CO fluxes taken from the literature using 
\begin{equation}
M_{H_{2}}=4.82\left( \frac{X}{3\times10^{20}}
 \right)L_{CO} 
\end{equation}
(see Tinney et al. 1990) where $L_{CO}$ is in $K km s^{-1} pc^{2}$. $I_{CO}$
values were taken from Sage, Salzer, Loose and Henkel (1992) and from
Taylor, Kobulnicky and Skillman (1998). 
$L_{CO}$ was calculated from the relation $L_{CO}=(\pi r^{2})I_{CO}$
where $r$ is the radius of the telescope beam on the galaxy in $pc$. 
Gas masses for the SLUGS sample were taken from Dunne et al. (2000).
For those objects whose $850\mu$m fluxes were 
obtained from Amure (priv. comm.) and Bianchi et al. (1998, 2000) we
took $S_{CO}$ values from Young et al. (1995).

We  investigated the dependence of the method on the
metallicity-dependence of the X-factor by trying two different
suggested relationships for the dependence of X on metallicity.
Arimoto, Sofue and Tsujimoto (1996) propose that X has a strong
dependence on metallicity.
\begin{equation}
log X^{\prime} = -1.0(12+log [O]/[H])+9.30
\end{equation}
where $X^{\prime}=N_{H_{2}}/I_{CO}\times10^{-20}H_{2}/(K km s^{-1})$.

Israel (2000) notes the probable dependence on ultra-violet ratiation
field, but also suggests a simplified approximate metallicity
dependence for  X, 
\begin{equation}
log X = 12.2-2.5log [O]/[H]
\end{equation}
We have used both of these relations to determine
$H_{2}$ masses for our sample. These are listed in Table 3. As an
additional test of the sensitivity of the method to the X-factor we 
have also used a ``standard'' conversion factor, $X=2.8 \times
10^{20} H_{2} cm^{-2}/[k km s^{-1}]$, from Bloemen et al. (1986)
derived from  observations of the
Galaxy. Table 3 also gives the dust masses calculated using equation 1.   

\section{Interpretation \& Discussion}
In figure 3 we have plotted the dust masses calculated by the two different 
methods against each other. We first performed a simple least-squares
fit on the data points.
All of the least squares solutions were consistent with a slope of
unity. This suggests that the properties of dust, in particular
$\kappa_{d}$ and $\varepsilon$, are similar for dwarf galaxies and large
galaxies like our own. On the assumption that this is true, the value
of $\kappa_{d}$ can be determined in the following way. If the value
of $\kappa_{d}$ we have initially assumed for plotting Fig. 3 is
correct, the points should lie on a line of slope unity passing
through the origin. By measuring the actual offset of the points from
the line, we can calculate a new value of $\kappa_{d}$.  
We obtain values of 0.07 $\pm$ 0.02, 
0.07$\pm$0.01  and 0.07$\pm$0.01 $m^{2} kg^{-1}$ 
using Arimoto, Israel and Galactic  $CO-H_{2}$ X-factors respectively.

The method clearly does not depend critically on the
metallicity-dependence of the CO-to-$H_{2}$ X-factor, presumably
because either the overall contribution of $H_{2}$ to the gas-mass is
in all cases relatively small, or, where the molecular content is
high, the X-factors do not vary greatly at these metallicities.  The main
uncertainty in the method is the basic assumption that the
fraction of metals in dust ($\varepsilon$) is a universal constant. If 
this is not correct, our estimate of this constant, from observations
of the interstellar medium in our own galaxy will not be applicable to 
other galaxies. However, Figure 3 is itself evidence that this is not
too bad an assumption. The scatter around the best fit lines is only
a factor of 2 in the ordinate. This scatter includes all observational errors,
as well as galaxy-to-galaxy variations in $\kappa_{d}$ and
$\varepsilon$. On the extreme assumption that all the scatter is
caused by variation in $\varepsilon$, we estimate that the
galaxy-to-galaxy variation in $\varepsilon$ can only be a factor of  $\sim  2$. If
this extreme assumption is correct, then our estimate of $\kappa$,
would have an error of a factor of
$\sim2$ for an individual galaxy.   

Our conclusion that the fraction of metals incorporated in dust is the 
same for giant and dwarf galaxies is in contradiction with the
conclusions of Lisenfeld \& Ferrara (1998), who found that metals are
less effectively incorporated in dust in dwarfs than in giant galaxies.
Figure 4 shows metallicity plotted against dust-to-gas mass ratio for
our sample and that of Lisenfeld \& Ferrara. 
The Lisenfeld \& Ferrara points suggest a non-linear
relationship between dust-to-gas ratio and metallicity.
However we have re-observed some of the objects from their sample with 
SCUBA and find that
their dust-to-gas mass ratios are increased when estimated from submm
rather than far-infrared flux. This is
expected since in their study, Lisenfeld et al. (1998) used the \emph{IRAS}
$60/100\mu$m flux ratio to determine a dust temperature and hence may have
underestimated the dust mass by missing cold dust at longer
wavelengths. The re-observed points are consistent with a slope of unity. 
Very recently Lisenfeld et al. (2001) have suggested that a low dust
mass is appropriate for NGC1569, whether such a model is applicable to all 
dwarf galaxies, or is unique to NGC1569 is not yet clear.
We have also plotted two of the galaxy dust evolution models of Edmunds
(2001) in figure 4. The solid line shows a model in which 
supernovae play an important role in dust production and the dashed
line a model in which dust is produced in some kinds of evolved low 
and intermediate mass
stars. Our current dataset is unable to make a distinction between
these two models but further submm observations at low metallicities
$(12 +log(O/H) < 7.8)$ should help demonstrate whether or not condensation
of dust grain cores in supernova ejecta is an important dust
source. Such observations
will be quite challenging as the number of objects suitable for observation at
$850\mu$m which have very low metallicities is small.  

In Figure 5 we have reproduced a plot from Alton (2001) that shows
estimates of
the dust mass absorption coefficient in the IR-Submm from a variety
of sources. Sopka et al. (1985) made observations of the
thermal emission from the dust envelopes of evolved stars. Rengarajan
(1984) reports results from observations of centrally heated infrared
sources deeply imbedded in molecular clouds. Boulanger et al. (1996)
studied the dust-to-gas correlation at high Galactic latitude using
COBE data. The Draine \& Lee (1984) points come from a model for
diffuse Galactic dust. Both the Hildebrand (1983) and Casey (1991)
points are from Galactic reflection nebulae. The Alton (2000)
measurement is from SCUBA observations of the spiral galaxy NGC891
while Bianchi et al. (1999) used COBE \& IRAS data. Finally, Agladze
et al. (1994) performed laboratory experiments on forsterite
$(Mg_{2}SiO_{4})$. The line
shows $\kappa \propto \nu^{2}$. Our data (open triangle) is consistent with the COBE
data of Boulanger et al. (1996) as well as that of Draine \& Lee
(1984), Hildebrand (1983) and Bianchi et al. (1999). 

\begin{figure}
\psfig{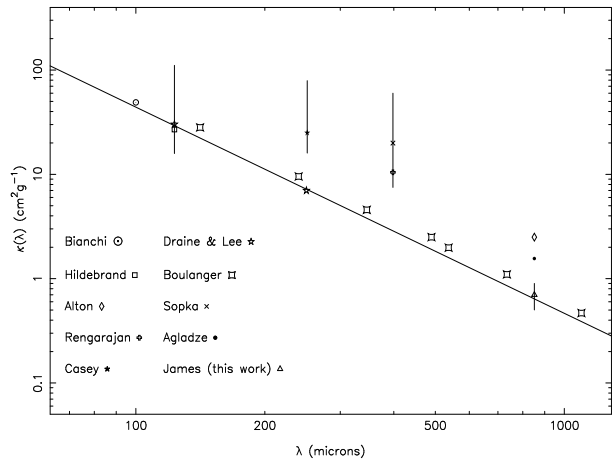}
\caption{The dust mass absorption coefficient in the submm (reproduced 
from Alton et al. (2001)). The data is from Hildebrand 1983, Casey 1991,
Rengarajan 1984, Sopka et al. 1985, Alton et al. 2000, Agladze et
al. 1994, Draine \& Lee 1984, Boulanger et al. 1996, Bianchi et
al. 1999 and this work. The data from Agladze et al. 1994 is for
amorphous dust grains at 20K. The line shows $\kappa \propto \nu^{2}$}
\end{figure}

\section{Conclusions}

We have presented a new method for determining the $850\mu$m dust mass
absorption coefficient,  
and obtain a value of  0.07$\pm$0.02 $m^{2} kg^{-1}$. 

Our method for determining this coefficient avoids the problems encountered by 
previous investigators (Hildebrand 1983, Bianchi 1999, Alton 2000). Using 
SCUBA observations at $850\mu$m we can probe 
the dust emission directly. We do not make any assumptions about the 
properties of the dust nor do we rely on similarity between dust in 
Galactic reflection nebulae and that in the general ISM.
Our only assumption (admittedly an important one) is that the fraction 
of metals incorporated in dust is a universal constant. This seems a
justified assumption however since the small scatter in Fig. 3
suggests that the fraction of metals in dust is the same for a wide range, 
from dwarf to high mass galaxies.

Having used both metallicity-dependent (Israel 2000; Arimoto et
al. 1996) and independent (Bloemen et al. 1986) $CO-H_{2}$ conversion
factors to derive a molecular gas mass we have concluded
that the metallicity-dependence of the $CO-H_{2}$
X-factor is not an important consideration in this derivation of
$\kappa_{d}$.   

The COBE data of Boulanger et al. (1996) are consistent with our value 
for $\kappa_{850}$ as are the data of Draine \& Lee, Bianchi and Hildebrand. 
The $\kappa_{850}$ value of Alton (2000) that was derived from SCUBA
observations of the edge on galaxy NGC891 is $\sim3$ times higher than 
ours. They do however admit that uncertainty in their technique means
that their quoted value for the dust emissivity could be in error by a 
factor of 2 either way. Agladaze et al. (1994) laboratory experiments yield a
$\kappa_{850}$ that is $\sim2$ times higher than ours, however they
use $\beta=1.3$ for wavelength extrapolations. Sopka et al. 1985 also
use $\beta=1.2$.
 
Unfortunately, the metallicity range of our sample does not extend
low enough to make a distinction between the dust formation models
proposed by Edmunds (2001). We intend to acquire $850\mu$m
data for the rest of the Lisenfeld \& Ferrara (1998) sample where
possible.  

\section*{Acknowledgments}
Guest User, Canadian Astronomy Data Centre, which is operated by the 
Dominion Astrophysical Observatory for the National Research Council of 
Canada's Herzberg Institute of Astrophysics. 

This research has made use of the NASA/IPAC Extragalactic Database
(NED) which is operated by the Jet Propulsion Laboratory, California 
Institute of Technology, under contract with the National Aeronautics and
Space Administration.

Stephen Eales thanks the Leverhulme Trust for the award of a research
fellowship. 

We thank the referee for helpful comments.

{}

\end{document}